\documentclass[aps,showpacs,amsmath,amssymb]{revtex4}

\usepackage{graphicx}
\usepackage{dcolumn}
\usepackage{bm}

\newcommand{\tf}{$T_f$\ }
\newcommand{\df}{$\Delta F^{\neq}$\ }
\newcommand{\pv}{$\phi$-value\ }
\newcommand{\pvs}{$\phi$-values\ }

\newcommand{\astr}{$\alpha^{\ast}$\ }
\def \kB{k_{\mbox{\tiny B}}}
\def \a{\alpha}
\def \Ff{F_{\mbox{\tiny F}}}
\def \Fu{F_{\mbox{\tiny U}}}
\def \Del{\Delta}
\def \Fdag{F^{\neq}}
\def \pF{p_{\mbox{\tiny FOLD}}}

\def \be{\begin{equation}}
\def \ee{\end{equation}}
\def \eps{\epsilon}

\begin{document}

\title{Three-body Interactions Improve the Prediction of Rate and
  Mechanism in Protein Folding Models } 
\author{M. R. Ejtehadi$^{\ddag\S}$}
\email{ejtehadi@physics.ubc.ca}
\author{S. P. Avall$^{\ddag}$}
\author{S. S. Plotkin$^{\ddag}$} 
\email{steve@physics.ubc.ca}

\affiliation{${}^\ddag$ Department of Physics and Astronomy, 
University of British Columbia, 
Vancouver, BC V6T-1Z1, 
Canada\\
${}^\S$ Department of Physics, 
Sharif University of Technology, 
Tehran 11365-9161, Iran
}

\date{\today}
\begin{abstract}
Here we study the effects of many-body
interactions on  rate and mechanism in protein folding, using the
results of molecular dynamics simulations on numerous coarse-grained
$C_\alpha$-model single-domain proteins. After adding
three-body interactions explicitly as a perturbation to a
G\={o}-like
Hamiltonian with native pair-wise interactions only, we have found 1) a
significantly increased correlation with experimental $\phi$-values
and folding rates,
2) a stronger correlation of folding rate with  contact
order,
matching the experimental range in rates when
the fraction of three-body energy in the native state is $\approx
20\%$, and 3) a considerably larger
amount of 3-body energy present in Chymotripsin inhibitor than other proteins studied. 
\end{abstract}

\maketitle

\baselineskip=0.6cm


Understanding the nature of the interactions that stabilize protein
structures and govern protein folding mechanisms is a fundamental
problem of molecular
biology~\cite{WolynesPG95:sci,Dobson98,FershtAR99:book,MirnyL01,DillKA97,Daggett03:nat},
with applications to structure and function prediction~\cite{MarcotteEM99,HaoM99:cosb,BonneauR01}
as well as rational enzyme  
design~\cite{BolonDN02}. 
Regarding folding mechanisms, protein folding has long been known to be a cooperative process, at
least for smaller single-domain
proteins~\cite{JacksonSE98}. Experimental scenarios that lack a first-order-like folding
barrier are rare~\cite{GruebeleM99}, often in contrast to simulation
results. There are other
discrepancies between simulation and experiment. For example,
while the experimental folding rates for a typical set of 18 2-state, single domain 
proteins (given in {\it Methods}) span $\sim 6$ orders of magnitude, 
simulations of coarse-grained
models of the same proteins have rates that vary by about a factor of
$100$, a discrepancy of 4 orders of magnitude.

How does one then quantify the sources of the barrier that controls
the folding rate? The folding  
barrier is the residual of an incomplete cancellation of large
and opposing energetic and entropic contributions, with the relative
smallness of the barrier allowing folding to occur on
biological
time-scales~\cite{Hao94,PlotkinSS02:quartrev12}. Among
the important energetic contributions that drive folding are
solvent-mediated hydrophobic forces~\cite{DillKA90r}, which are known
to be weaker on short length scales, or low concentrations of apolar
side-chains~\cite{LumK99}- a scenario likely to be present when the
protein is unfolded. Hence the solvent-averaged potential governing
folding almost certainly contains a non-additive, many-body
component.
The folding free energy barrier increases as
the non-additivity of interactions is
increased~\cite{PlotkinSS97,Doyle97,ChanHS00:pro},
due to the decreased energetic correlation between the native
conformation and conformations that may be geometrically similar to
it. 

Experimental $\phi$-values give a measure of the strength of native
interactions involving a particular amino acid (residue) in the
transition state~\cite{FershtAR92}, thus quantifying a residue's
importance in folding. However the $\phi$-values obtained from
simulations of coarse-grained protein models generally do not
correlate well with the experimentally determined
values. Model proteins are coarse-grained on the
belief that a reduced number of degrees of freedom can capture the
essentials of the folding
process~\cite{OnuchicJN95:pnas,SheaJE01,MirnyL01}, however the less
than ideal agreement with experimentally observed rates and mechanisms
leads one to consider alternate forms for the coarse-grained
Hamiltonian or energy function, as well as to consider more detailed
all-atom models~\cite{DaggetV94,YoungWS96,SnowCD02} which may contain
explicit solvent as well~\cite{BoczkoEM95,DuanY98,KazmirskiSL01,Daggett03:nat,SnowCD02,GarciaAE03:pnas}.

But it is also clear that coarse-grained simulations allow a study
of microscopic dynamics that would not be possible by all-atom models
with present-day computing power. 
Because we cannot yet fully analyze the statistics of folding trajectories in all-atom
models, coarse-grained simulational models such as off-lattice $C_\alpha$
models~\cite{GuoZ95,ZhouY99,ClementiC00:jmb,VendruscoloM01,MirnyL01,KogaN01,SheaJE01}
have been essential in elucidating protein-folding mechanisms. 

We could then take the following
approach:  postulate a given feature thought to be present 
in the system
and ask to what extent this feature, such as many-body
potentials, must be present in the Hamiltonian of a coarse-grained
model for best agreement with existing experimental data on
protein folding rates and mechanisms.


\section{Materials and Methods}

\subsection{Simulation Model}
Eighteen two-state folding proteins with known native structures (PDB codes 
1AEY, 1APS, 
1FKB, 1HRC,  1MJC, 1NYF,  1SRL, 1UBQ, 1YCC,
2AIT, 2CI2, 1PTL, 2U1A, 1AB7, 1CSP, 1LMB, 1NMG, 1SHG ) were
selected for coarse-grained simulations.
For all proteins except the last 5 above, rate data
was available at various denaturant concentrations. These were then
used for further analysis at the stability of the  
transition midpoint. 

The simulated proteins consist of a chain of connected beads, with
each bead representing the position of the $C_\alpha$ atom in the
corresponding amino acid.  The off-lattice $C_\alpha$ G\={o} model has been
described in detail
previously~\cite{GuoZ95,ClementiC00:pnas,SheaJE01,KogaN01}. 
The Hamiltonian has local
and non-local parts: Bond, angle and dihedral angle potentials
constitute local interactions. In the putative G\=o model, pair
contacts between residues in spatial proximity in the native structure
constitute non-local interactions.  Non-native interactions are
treated by a sterically repulsive pair-potential only. 
Heavy atoms within a cut-off distance of $r_c = 4.8$ \AA\ in
the native structure obtained from the PDB file are associated with a
Lennard-Jones-like 10-12 potential of depth $\epsilon_2 = -\kB T$ and a position
of the minimum equal to the distance of the $C_\alpha$ atoms in the
native structure. Let there be $N_2$ pair-contacts of energy
$\epsilon_2$ in the native PDB structure. Then in an arbitrary conformation
there are $Q N_2$ contacts with energy $E_2 \approx
\epsilon_2\,Q \,N_2$, with $Q$ the fraction of native pair contacts
(we account for the continuum nature of the Lennard-Jones potentials).

We let triples with heavy atoms within a cutoff distance of $4.8$
\AA \  in the native structure have an energy
$\epsilon_3$. 
For a given protein there will then be $N_3$ 3-body contacts present
in the PDB native structure, with total $3$-body energy 
$\epsilon_3 N_3$.
An arbitrary structure then has a 3-body contribution
to the energy of $E_3 \equiv 
\epsilon_3\,Q_3\,N_3$,  where $Q_3$ is the fraction of
native triples present in that conformation.
Three-body
interactions are again G\=o-like; the remaining bond, angle, dihedral,
and non-native interaction energies are all unchanged.

When both pair-wise and 3-body interactions are present, 
the native non-local part of the energy becomes:
\begin{equation} 
E_{\mbox{\tiny{NL}}}(\a) = (1-\alpha) E_2 + \alpha E_3.
\label{eq:ENL3body}
\end{equation} 
The free parameter $\alpha$ ($0\leq \a \leq 1$) controls the relative contribution
of two- and three-body interactions.
The energy per triple is assigned as $\epsilon_3 = \epsilon_2 \,
N_2/N_3$, to preserve overall native stability.

Dense sampling is obtained from long simulations with a purely 2-body G\={o}
Hamiltonian at the transition mid-point (e.g. for CI2 the
simulation time corresponds to about $3$ seconds, as
determined from the number of folding and unfolding events).
From histograms of the number of states at a given fraction of native
contacts $Q$, the free energy $F(Q)$ can be constructed. All simulated
free energy profiles displayed a single dominant barrier.
All proteins are considered at their transition mid-points only, where
the unfolded and folded free energies are equal: $\Fu = \Ff$ 
(figure~\ref{fig:df_alpha}~A).

Three-body energies are treated as a perturbation
on the Hamiltonian. The new free energy is given by the exact expression:
\begin{equation} 
\frac{F(Q,\a)}{k_{\rm B} T}  = - \ln \frac{\sum_i \mbox{e}^{- \Delta
E_i(\a)/k_{\rm B} T}\, \Del(Q^{(i)}, Q)}{\sum_i e^{- \Delta E_i(\a)/k_{\rm B}
    T}},
\label{eq:zwanzig}
\end{equation}
where the sum is on all sampled conformations $i$, 
$\Del(Q^{(i)}, Q)$ is a delta function that selects only those
states where $Q^{(i)} = Q$, and 
$\Del E (\a) = E_{\mbox{\tiny{NL}}}(\a) - E_2$.

\subsection{Calculated $\phi$-values}

Simulated kinetic \pvs are 
given by~\cite{Onuchic96}:
\begin{equation}
\label{eq:phi2b}
\phi_i = \frac{\langle n_i \rangle_{\neq} - \langle n_i \rangle_U}
{\langle n_i \rangle_F -\langle n_i \rangle_U},
\end{equation}
where $\langle n_i \rangle$ is the thermal mean value of number of
contacts for residue $i$, and the $\neq$, $U$ and $F$ subscripts refer to
the transition state, unfolded state and folded state ensembles
respectively. 

We first compare simulated and experimental \pvs using  the thermal
transition state ensemble (TTSE)  around the free energy
barrier peak, i.e.  $\left|F
-F^{\neq}\right|/\Del \Fdag \leq 0.2$ was used to define a width
$\Del Q$ of the barrier peak
(shaded in figure~\ref{fig:df_alpha}~A). Conformations within this range were
taken to be the TTSE, and were used to calculate $\phi$ values from
equation~\ref{eq:phi2b}.  
The validity of the TTSE was checked for CI2
and SH3 with a comparison of \pvs using the kinetic transition state
ensemble (KTSE), selected as having a folding probability $\pF$ of roughly
$1/2$~\cite{DuR98:jcp}.  Conformations in the
TTSE were used as initial conditions for 100 simulations which were
terminated when the protein folded or unfolded. Those conformations
that had a $\pF$ within $0.5 \pm 1/\sqrt{100}$ were taken as the KTSE.
For CI2 (SH3) we found $315$ ($283$) KTSE configurations from a total
of $2359$ ($2078$) TTSE
configurations.

Other reaction coordinates were helpful in determining the kinetic
transition state ensemble by constructing multi-dimensional reaction
surfaces. To this end we found a contact-order 
weighted variant of $Q$ to be useful, which for any configuration
$\nu$ is given by:
\be
Q_{\mbox{\tiny{CO}}}^{\nu} = \frac{\sum_{i<j} \left| i- j\right|
  \Del_{ij}^{\nu} \Del_{ij}^{N}}{\sum_{i<j} \left| i- j\right|
  \Del_{ij}^{N}}
\label{QCO}
\ee
where the sum is over all $C_\alpha$  atoms, and $\Del_{ij}^{\nu}$ and $\Del_{ij}^{N}$
are unity if residues $i$ and $j$ are in contact in conformations
$\nu$ and the native structure respectively, otherwise they are zero.

We determined
$\phi$-values in the  presence of
three-body interactions analogously to eq.~(\ref{eq:phi2b}). 
Under some simplifying assumptions (e.g. requiring that the
\pv that is independent of the perturbation energies):
\begin{widetext}
\begin{equation}
\label{eq:phi_alpha2}
\phi_i^{(\alpha)} = 
	\frac{(1-\alpha)\, (\langle n_i \rangle_{\neq}^{(\alpha)} 
			- \langle n_i \rangle_U^{(\alpha)})\,N_3 
		+ \alpha\, (\langle m_i \rangle_{\neq}^{(\alpha)} 
			- \langle m_i \rangle_U^{(\alpha)})\,N_2 } 
	     {(1-\alpha)\, (\langle n_i \rangle_F^{(\alpha)}
			- \langle n_i \rangle_U^{(\alpha)})\,N_3 
		+ \alpha\, (\langle m_i \rangle_F^{(\alpha)} 
			- \langle m_i \rangle_U^{(\alpha)})\,N_2 }.
\end{equation}
\end{widetext}
Here  $m_i$ is the number of three body interactions in which monomer
$i$ is involved, and superscript $(\alpha)$ indicates averaging the
ensembles ($\neq$, $U$, $F$)  in the presence of  3-body energy. 
When $\a \rightarrow 0$, (\ref{eq:phi_alpha2}) reduces to (\ref{eq:phi2b}).

\subsection{Miyazawa-Jernigan-based Models}

The effect of heterogenity in the model was also studied by
interpolating  between the G\={o} model 
and the Miyazawa-Jernigan (MJ) models by varying the
free parameter $\a$ between zero (Homogeneous G\={o} model) and
unity (MJ model). The contact
energy for any pair of residues (not necessarily native) is then:
\be
\eps_{ij} = (1-\a) \epsilon_2 + \a \epsilon_{ij}^{\rm
MJ} \: ,
\label{eq:mj}
\ee
where $\epsilon_2 $ is as above, and
$\epsilon_{ij}^{\rm MJ}$ was proportional to the MJ interaction
energy~\cite{MiyazawaS96} between the residue types of
$i$ and $j$, scaled by a factor to ensure  the energy of the native
structure is $\a$-independent. An interpolation between a uniform
G\={o} model and a heterogeneous G\={o} model with native contact energies
given by MJ parameters was also considered. 

\subsection{Contact Order and Statistical Significance}
Absolute contact order is the average sequence separation between residues having native
contacts~\cite{Plaxco98}: $aCO =  M^{-1} \sum_{i>j} |i-j|$, where $M$ is
the total number of native contacts. Relative contact order is scaled
again by chain length $N$: $rCO = aCO/N$.

Statistical significance or $P$-value is the probability to
achieve a given correlation coefficient, $r$, assuming random data: $P =
\mbox{erf}(|r|\sqrt{N/2})$. 
Small data sets almost always have fairly large $P$, even if $r$ is
large. Large data sets may still have small $P$ even if the
correlation is weak, which would still indicate a systematic effect.


\section{RESULTS}
\subsection{Protein folding rates}
 
Here we considered the effect of introducing a three-body potential to
an off-lattice two-body G\=o model studied
previously~\cite{ClementiC00:pnas,SheaJE99,KogaN01}.  Eighteen mentioned
single-domain proteins that are known to fold by a two-state mechanism
were selected, and coarse-grained so that each amino
acid corresponds to a bead at the position of the $C_\alpha$ atom.
Long simulations at the folding temperature \tf for a subset of the
proteins showed a single exponential distribution of first passage
times: $P(\tau) \sim \exp(-\kappa t)$. For these proteins the
simulated log folding rate, $\mbox{log}(\kappa)$, correlated very strongly
(r=0.997) with the free energy barrier height \df,
indicating that \df was an accurate predictor of the rate for the
simulated G\={o} models. We subsequently assume this proportionality
between \df and $-\mbox{log}(\kappa)$ for all simulated proteins, referring
to $\exp(-\Delta F^{\neq}/\kB T)$ as the ``effective rate''.

The above mentioned discrepancy between the effective
protein rates for our data set and the
experimentally determined rates for the same proteins motivates an
investigation of the effect of many-body interactions on rates. When a
portion of the total energy is attributable to many-body interactions,
energetic gain is not achieved until a larger amount of native
structure is present, with a correspondingly larger entropic
cost. Several polymer loops must be simultaneously closed during
folding to receive energetic gain. This effect enhances the dependence of
rate on contact order, increasing the range over which rates
vary.

By attributing a fraction $\alpha$ of the native energy to triples in
the native structure,  we studied the effects of three-body
interactions by varying this single parameter (see Methods). 
The effects on the free energetic potential surface for several
proteins are shown in figure~\ref{fig:df_alpha}.

As the fraction of 3-body energy is increased, the correlation of
the simulated effective rates with both absolute and relative contact
order increases (figure~\ref{fig:df_rco} a,b).
This effect has also been seen in lattice protein
models~\cite{KayaH03:prots2,JewettAI03}.  
We can also quantify how much 3-body energy, at the residue level, reproduces the
experimental dispersion in rates for single-domain proteins. The
simulated effective rates span 6 orders of magnitude when
approximately $20\%$ of the energy in the native state of the
coarse-grained protein is due to 3-body interactions.

Rates simulated with a 2-body Hamiltonian do not correlate
significantly with
experimentally determined rates at $25^o C$ (figure~\ref{fig:df_rco}~C). 
We can remove the effects due to variations in stability and reflect
the conditions in the simulations by taking instead the 
rate data at the various transition midpoints (after the addition of
GdHCl). We then 
found the correlation significantly increased to $r=0.64$, $p=0.018$. 
Adding 3 body energy in the simulations increases the correlation with 
the experimental
rates (at the transition midpoints) still further, with 
the best correlation  achieved when $\alpha= 10\%$
(see figure~\ref{fig:df_rco}d).

These results
strongly suggest that 1) stability is an important 
determinant of 
folding rate, 2) many-body energy is present in the energy functions of
real proteins,
and 3) G\={o} or G\={o}-like models (which ignore non-native
interactions) can predict experimental rates, illustrating the minor
importance of non-native interactions in governing folding barriers.

The correlation of log rates with $rCO$ also improves as $\alpha$ is
increased from zero, however the correlations are modest,
increasing from ($r= 0.29$, $P = 0.24$) at $\a =0$ to a best
correlation of ($r=-0.44$, $P=0.08$) at $\alpha = 10\%$ (data not shown).

\subsection{Testing pair interaction matrices}

The correlation between experimental and simulational \pvs for a
2-body Hamiltonian ($r_0$, $P_0$) was typically not statistically significant (see
table~I), with the exception of SH3.  
Rank ordered measures of correlation such as Kendall's tau, which are
insensitive to the precise values of the data, generally do not
improve the agreement~(table~II). 
We also checked whether simulations with a 2-body Hamiltonian could
accurately 
predict residues that had higher-$\phi$
values. This was done by 
weighting the statistical averaging in the
correlation coefficient by the experimental \pv itself as a Jacobian
factor. Implementing this recipe did not substantially increase the
correlation coefficient, and in fact decreased it in the cases of
AcP and CI2~(table~I). Similar results were obtained by implementing a
simple cut-off imposing a lower bound for relevant experimental \pvs
(data not shown). 

The experimental data can be used to test energy functions
characterizing pair-interactions at the amino acid level, such as the
Miyazawa-Jernigan (MJ) matrix~\cite{MiyazawaS96}.  We investigated
whether MJ interaction parameters improved the simulational
predictions of $\phi$-values, by interpolating between a homogeneous
G\={o} model and a model with pair interactions (between all residues)
governed by MJ
parameters (see equation~(\ref{eq:mj})).  We also
interpolated between a homogeneous G\={o} model and 
a heterogeneous G\={o} model with native interaction parameters
determined from the MJ matrix.

Results are shown for two proteins in
figure~\ref{fig:phi_corr_alpha}. For CI2 and SH3, no improvement in the
correlation with experimental data was seen by implementing this
procedure. Table~I shows the results for the comparison
between experimental \pv data and \pvs obtained from a pairwise MJ
Hamiltonian. In general if correlations increased by interpolating
toward MJ parameters they did so only
modestly- only in the case of protein~L did the improvement reach statistical
significance ($P=1\%$, see table~I). 

To check of the validity of the recipe of interpolating toward MJ
parameters, we compared the largest improvement in correlation $(r_{\alpha^*}
- r_o)$ with the value $\alpha^*$ of three body energy required to achieve that correlation.
This tests whether the poorness of the original correlation was due to
the absence of MJ coupling energies.  We found that $(r_{\alpha^*}
- r_o)$ itself correlated well with $\alpha^*$, however the
statistical significance was not particularly strong, and the slope
measuring the degree of improvement was not particularly high (see
figure~\ref{fig:drvsa}).

\subsection{Testing three-body interactions}

The experimental data can also be used as a benchmark to test what
amount of 3-body
energy in the Hamiltonian of the coarse-grained model gives best
agreement with experimental \pvs. 
We examined this question 
for the 5 proteins in table~I, by
measuring the correlation between the experimentally obtained
$\phi$-values, and $\phi$-values of the same residues determined from
simulations, with conditions ranging from between a pair-wise interacting
G\={o} model
protein, and one governed exclusively by 3-body interactions at the
residue level (see methods).

As the strength of 3-body interactions increased from zero,
the correlation coefficient also increased, for all proteins
studied~(see fig.~\ref{fig:phi_corr_alpha} and table~I).  An exceptional case was
SH3, which showed only a modest increase in correlation for the
kinetically determined transition state ensemble, and no
increase for the thermal transition state ensemble. The fraction \astr
of native 3-body energy that gave best agreement with experimental
data varied from protein to protein, but correlated strongly with the
increase in agreement with experimental data (see table~I).  
That is, the improvement in
correlation $(r_{\alpha^*} - r_o)$ itself correlated very
strongly with \astr ($r=0.97$, $P=0.005$), further supporting the notion
that the poorness of the 
original agreement was due at least in part to the absence of
many-body forces.

For a protein such as CI2 with large fraction of 3-body energy, the
transition states in the presence of 3-body interactions is
significantly different than the 2-body transition
state. 
For CI2, the root mean square
distance (RMSD) between all $315$ structures in 
the kinetic transition state ensemble (KTSE) was found for both the 2-body  and 2+3-body (at $\a^*$)
cases.
Shown in figure~\ref{fig:TS}A, B is the ``most representative'' transition state
structure for the 2-body  and 2+3-body cases respectively, defined as
having the minimal Boltzmann-weighted RMSD (minimum over
structure $i$ of $\sum_j \, p_j (\mbox{RMSD})_{ij}$) to all others in the KTSE.
The 2-body case shows more overall secondary structure, in particular
more $\alpha$-helix, but less $\beta$-sheet. The $Q$, 
$Q_{\mbox{\tiny{CO}}}$ (see methods), and $R$ (RMSD from the native structure) values
for the structures in figure~\ref{fig:TS}A,B are 
$Q^{\mbox{\tiny{(A)}}} = 0.54$, $Q^{\mbox{\tiny{(B)}}} = 0.49$, 
$Q_{\mbox{\tiny{CO}}}^{\mbox{\tiny{(A)}}} = 0.41$,
$Q_{\mbox{\tiny{CO}}}^{\mbox{\tiny{(B)}}} = 0.29$, and
$R^{\mbox{\tiny{(A)}}} = 5.5$ \AA, $R^{\mbox{\tiny{(B)}}} = 11$ \AA. 
This indicates that the 2+3-body transition state is less structured
than the pure 2-body transition state. 
However, kinetically they are
about the same distance from the native structure, with  $\pF$ values 
$\pF^{\mbox{\tiny{(A)}}} = 0.55$, $\pF^{\mbox{\tiny{(B)}}} =
0.53$~\cite{note:pfold}.
They have a RMSD of
$7.8$ \AA \ between them, so they are structurally distinct from each
other.
The average RMSD values from the native for the top 4 transition state
structures for the 2-body and (2+3)-body cases are
$\overline{R}^{\mbox{\tiny{(2)}}} = 6.3$ \AA,  and 
$\overline{R}^{\mbox{\tiny{(2+3)}}} = 8.5$ \AA,  again confirming less
native structure in the more accurate transition state containing
3-body interactions. 
Interestingly, the high-$\phi$ residue 34 has more local secondary
structure in the pure 2-body case than at $\a^*$. It also has no triples in the native
state. Its high \pv in the presence of 3-body interactions is the
result of correlations with other triples made in the transition
state.

The procedure of adding 3-body interactions was repeated considering
only residues in the hydrophobic core of native structure, in this
case buried with
less than $\approx 30\%$ accessible surface area using the Swiss PDB
algorithm. (http://www.expasy.org/spdbv).
We saw qualitatively the same effect, but the change in correlation coefficient
was less pronounced, increasing to about $0.42$ for CI2 for example. 
This implies that 
coarse-grained model proteins  with effective solvent-averaged
interactions have many-body interactions involving 
residues on the surface as well.

\section{Discussion}

The above results suggest that many-body interactions can play a
significant role in governing the folding mechanisms of 2-state
proteins when described at the residue level. 
This seems quite evident upon comparing the statistical significance
columns in table~I or table~II for the pure 2-Body Hamiltonian and the 2+3-body
Hamiltonian at $\a^*$.
In essentially all cases, many-body interactions helped to establish
consistency with protein folding experiments. 
Some proteins showed
dramatic improvement, others mild improvement, so proteins may be
additionally classified through this effect. 
The value
of \astr may be used as an indication of the importance of many-body interactions
in governing the folding mechanism for a given protein, as for example
the proteins are ranked in tables~I and~II.

Experimental rates vary by about 4 orders of magnitude more than rates
obtained from coarse-grained models using 2-body Hamiltonians.
However a modest 3-body component to native stability (about 20\% on average) 
was sufficient to reproduce the experimental variability in
folding rates. It is an open question as to how large the many-body
component might be in finer-scale and all-atom models of proteins.
Ab initio studies of interaction energies and reconfiguration barriers in water clusters
suggest they can be quite significant~\cite{MiletA99}.

For FKBP, protein~L, and CI2 the correlation between experimental
and simulational $\phi$ values goes from insignificant to significant
as 3-body interactions are added. 
In the case of CI2, the agreement between simulations with a 2-body energy function and
experimental data was the poorest of the proteins
studied, the fraction of 3-body energy at best agreement was the
largest, and the improvement in correlation coefficient  the most
dramatic. In the case of SH3 on the other hand, the folding mechanism appears to be
governed more by topology than by energetic considerations.
In some sense this is an exception that proves the rule, since
previous evidence supported a folding mechanism dominated by
topological considerations~\cite{RiddleDS99,MartinezJC99}.

Interestingly, muscle
acylphosphatase had the poorest improvement in mechanism
prediction by adding 3-body interactions, as measured by the
correlation coefficient. Its original  
$\phi$-correlation for a 2-body G\={o} model was the second poorest
after CI2. It also required the largest amount of Miyazawa-Jernigan
interactions for best agreement with experimental \pvs, but still
correlated poorly even at best agreement. Intriguingly it is also the
slowest known 2-state folder at present, yet a good 2-state folder
with no intermediates~\cite{ChitiF99}. The slow folding is likely due
to large contact order however, and it would be interesting in the
future to apply
the 3-body recipe to a topologically similar but faster folding
protein such as human procarboxypeptidase A2.
On the other hand, the improvement for AcP as measured by Kendall's tau does in fact become
statistically significant, and suggests a large 3-body
component. We are inclined to take this more robust measure of
statistical significance more seriously. 
The discrepancy of $r$ and $\tau$ indicates some large outliers in
\pvs, likely due to variations in native stabilizing interactions, 
which may exist for functional reasons. These fluctuations in native
interaction strength are
not captured by the uniform G\={o} model and 2+3-body models.

The largest improvement in correlation $(r_{\alpha^*}
- r_o)$ with the value of interpolation parameter $\alpha^*$ required to
achieve that correlation was used as a measure to test the validity of
the 3-body and Miyazawa-Jernigan interpolation recipes. 
The results for the 3-body interpolation recipe showed a strong
statistically significant correlation with large slope indicating
large rate of improvement. 
The results for the heterogeneous MJ G\={o}
model also showed improvement, however with smaller slope and smaller
statistical significance. It is
noteworthy that for the case where CI2, where the 3-body
recipe does the best, the MJ recipe failed to improve the agreement
with experiment.

For CI2, the transition state in the
presence of 3-body interactions shows less overall native structure
than the purely 2-body transition state, in spite of the better
agreement with experimental \pvs for the 3-body case. However 
it is not clear that this will be a general rule. 
In both
cases the transition state consists largely of a disordered form of
the native topology, sufficiently disordered to be kinetically
balanced between the folded and unfolded states.

The low levels of agreement between experiment and simulation for
2-body Hamiltonians told a somewhat cautionary tale. While a large
body of evidence leaves little doubt as to the importance of native topology in
governing folding mechanism, these results should serve to show
that realistic aspects of the energy function, such a many-body
component to native stability, should not be ignored.

\section{Acknowledgments} 
S. S. P. acknowledges support from the
Natural Sciences and Engineering Research Council and the Canada
Research Chairs program. We thank Cecilia Clementi and Baris Oztop for
helpful discussions. 

\newpage
\bibliographystyle{PNAS}
\bibliography{steve}

\begin{thebibliography}{10}
\bibitem{note:pfold}
Note however that $\pF$ values are only calculated in the presence of the
  2-body Hamiltonian, so the KTSE is only approximate in the presence of a 2+3
  body Hamiltonian .
\end{thebibliography}
\newpage

\begin{table*}
\label{tab:1}
\caption{
{\bf Two-body and Three-body characterization of proteins studied}
}
\begin{ruledtabular}
\begin{tabular}{l|rl|rrl|rrl|rrl}
& \multicolumn{2}{c}{G\=o  MODEL\footnotemark[2]}
& \multicolumn{3}{c}{MJ MODEL}
&  \multicolumn{3}{c}{MJ-G\=o MODEL}
& \multicolumn{3}{c}{3-BODY MODEL}
  \\
Proteins (PDB) \footnotemark[1] & $r_0$ & $P_0$ & $\alpha^*$ \footnotemark[3]& $r_{\alpha^*}$  \footnotemark[4]& $P_{\alpha^*}$  \footnotemark[4]&
$\alpha^*$ & $r_{\alpha^*}$ & $P_{\alpha^*}$ &
$\alpha^*$ & $r_{\alpha^*}$ & $P_{\alpha^*}$  \\ \hline

SH3 (1SRL)  &
0.58 \footnotemark[5]& 0.0003  & 0\% & 0.59 &  0.0003  & 5\% & 0.59 &  0.0002 & 5\% & 0.60 \footnotemark[5] &  0.0001  \\

FKBP (1FKB) &  0.32 & 0.17 &  10\% & 0.41 & 0.07 & 20\% & 0.38 & 0.1  & 10\% & 0.43 & 0.057  \\

AcP (1APS)  &   0.12 & 0.58  &  50\% & 0.35 & 0.1  & 30\% & 0.30 & 0.16 & 15\% & 0.32& 0.14  \\

Protein L (2PTL) &  0.18& 0.25 &  20\% & 0.38 & 0.01 &  30\% & 0.38 & 0.01 & 15\% & 0.53 & 0.00027 \\

CI2 (2CI2) &   -0.10 \footnotemark[5] & 0.56\footnotemark[6] &  0\% &
-0.017  &  0.92 & 0\% & -0.017  &  0.92 & 35\% & 0.57 \footnotemark[5] &
0.0004   
\end{tabular}
\begin{tabular}{l|cccc|cccc|rl}
&&&&& \multicolumn{4}{c}{3-BODY MODEL}& 
\multicolumn{2}{c}{HIGH-$\phi$ WEIGHTING} \\

Continue & $N$ \footnotemark[7] &
$N_2$\footnotemark[8] &
$N_3$\footnotemark[9]  &
$n$\footnotemark[10] & 
$\alpha^*$ &
$\Delta F^{\neq}_{\alpha^*}$\footnotemark[11]&
$\frac{\Delta F^{\neq}_{\alpha^*}}{\Delta
F^{\neq}_0}$\footnotemark[12]&
$ \frac{E_{\rm 3B}^{\rm \neq}}{E_{\rm tot}^{\rm \neq}}$\footnotemark[13] &
$\widetilde{r}_{0}$\footnotemark[14]&
$\widetilde{P}_{0}$\footnotemark[14]
\\ \hline

SH3 (1SRL) &
56 & 128 & 32 & 35 & 5\% & 3.8 $\pm$ 0.2 &1.4 & 2.6\%\footnotemark[5] &
0.65\footnotemark[5] & $2.7\times 10^{-5}$ \\

FKBP (1FKB) &
107 & 299 & 111 & 20 &10\% & 10 $\pm$ 0.8 &1.5 &5.5\% &
0.37 & 0.10 \\

AcP (1APS) &
98 & 257 & 97 & 23 & 15\% &14 $\pm$ 2.0 &2.2 &8.9\% & 
-0.02 & 0.91 \\

Protein L (2PTL) & 62 & 126 & 30 & 41 &15\% & 6.2 $\pm$ 0.5 &2.8 &3.3\% &
0.26 & 0.10 \\

CI2 (2CI2)) &  65 & 148 & 54 & 35 & 35\% & 17 $\pm$ 3.5 & 3.4 &
13\%\footnotemark[5] &
-0.43\footnotemark[5]& 0.01

\end{tabular}
\footnotetext[1]{Sources for experimental $\phi$-value data:
src-SH3 domain~\cite{RiddleDS99},FKBP~\cite{FultonKF99},
AcP~\cite{ChitiF99}, CI2~\cite{Itzhaki95}, 
protein~L~\cite{KimDE00}. }
\footnotetext[2]{Correlation coefficient and statistical significance
 between experiments and simulations of a pair-wise interacting
 G\={o} model.}
\footnotetext[3]{$\a^*$ is in general the value of the interpolation
  parameter that gives best agreement with experimental data
for corresponding model. For the MJ models eq.~(\ref{eq:mj}) is used,
for the 3-body models eq.~(\ref{eq:ENL3body}) is used. }
\footnotetext[4]{ $r_{\alpha^*}$ and $P_{\alpha^*}$ are the correlation coefficient and
 statistical significance respectively, at best agreement for the corresponding model.}
\footnotetext[5]{Kinetic transition state (KTSE) has been used.}
\footnotetext[6]{We allow for the possibility of anti-cooperativity in
  proteins, and hence ascribe statistical significance to negative correlations. Thus
  P-values here are the 2-sided statistical significance.}
\footnotetext[7]{Chain length.}
\footnotetext[8]{Number of native pair contacts.}
\footnotetext[9]{Number of native triples.}
\footnotetext[10]{Number of \pv data points used in the comparison.} 
\footnotetext[11]{Barrier height in $\kB T$ at \astr.}
\footnotetext[12]{Ratio of the free energy barriers when
 $\alpha=$\astr and $\alpha=0$.}
\footnotetext[13]{Fraction of 3-body energy in the
 transition state ensemble at \astr.}
\footnotetext[14]{Correlation coefficient and
statistical significance including a Jacobian factor weighting each term in
the correlation function by the experimental \pv itself, i.e. averages
are calculated as $\left< A\right> = 
(\sum_1^n \phi_i^{\rm exp} A_i)/(\sum_1^n \phi_i^{\rm exp})$  
where $n$ is the number of data points. 
This is a recipe simply to stress the importance of the agreement between large \pvs. }
\end{ruledtabular}
\end{table*} 

\begin{table}
\label{tab:2}
\caption{
{\bf Kendall's $\tau$ and Statistical significance between experiment
  and simulation}
}
\begin{ruledtabular}
\begin{tabular}{l|rl|rrl}
& \multicolumn{2}{c}{G\=o  MODEL\footnotemark[1]}
& \multicolumn{3}{c}{3-BODY MODEL\footnotemark[2] }
  \\
Proteins (PDB) & $\tau_0$ & $P_0$ & 
$\alpha^*$ & $\tau_{\alpha^*}$ & $P_{\alpha^*}$  \\ \hline

SH3 (1SRL)  &
0.42 \footnotemark[3]& 0.00044  &  0\% & 0.42 \footnotemark[3] &  0.00044  \\

FKBP (1FKB) &  0.27 & 0.10 &  10\% & 0.31 & 0.055  \\

Protein L (2PTL) &  0.14& 0.19 & 20\% & 0.36 & 0.00069 \\

AcP (1APS)  &   0.14 & 0.37  &  25\% & 0.33 & 0.027  \\

CI2 (2CI2) &   0.042 \footnotemark[3] & 0.72 & 35\% & 0.40 \footnotemark[3] &
0.0008   
\end{tabular}
\footnotetext[1]{Kendall's tau measure of ranked correlation and
 statistical significance ($P(|\tau'| \geq |\tau|)$) of tau value,
 between experiments and simulations of a pair-wise interacting 
 G\={o} model.}
\footnotetext[2]{$\a^*$ is the value of the interpolation
  parameter that gives best agreement with experimental data
for a 2+3-body Hamiltonian as in
 eq.~(\ref{eq:ENL3body}). $\tau_{\alpha^*}$ and $P_{\alpha^*}$ are
 Kendall's $\tau$ and 
 statistical significance respectively, at best
 agreement for the 2+3-body model.}
\footnotetext[3]{Kinetic transition state (KTSE) has been used.}
\end{ruledtabular}
\end{table} 

\newpage
.\newpage
{\large {\bf FIGURE CAPTIONS}}

\begin{figure}
\includegraphics[width=0.8\linewidth]{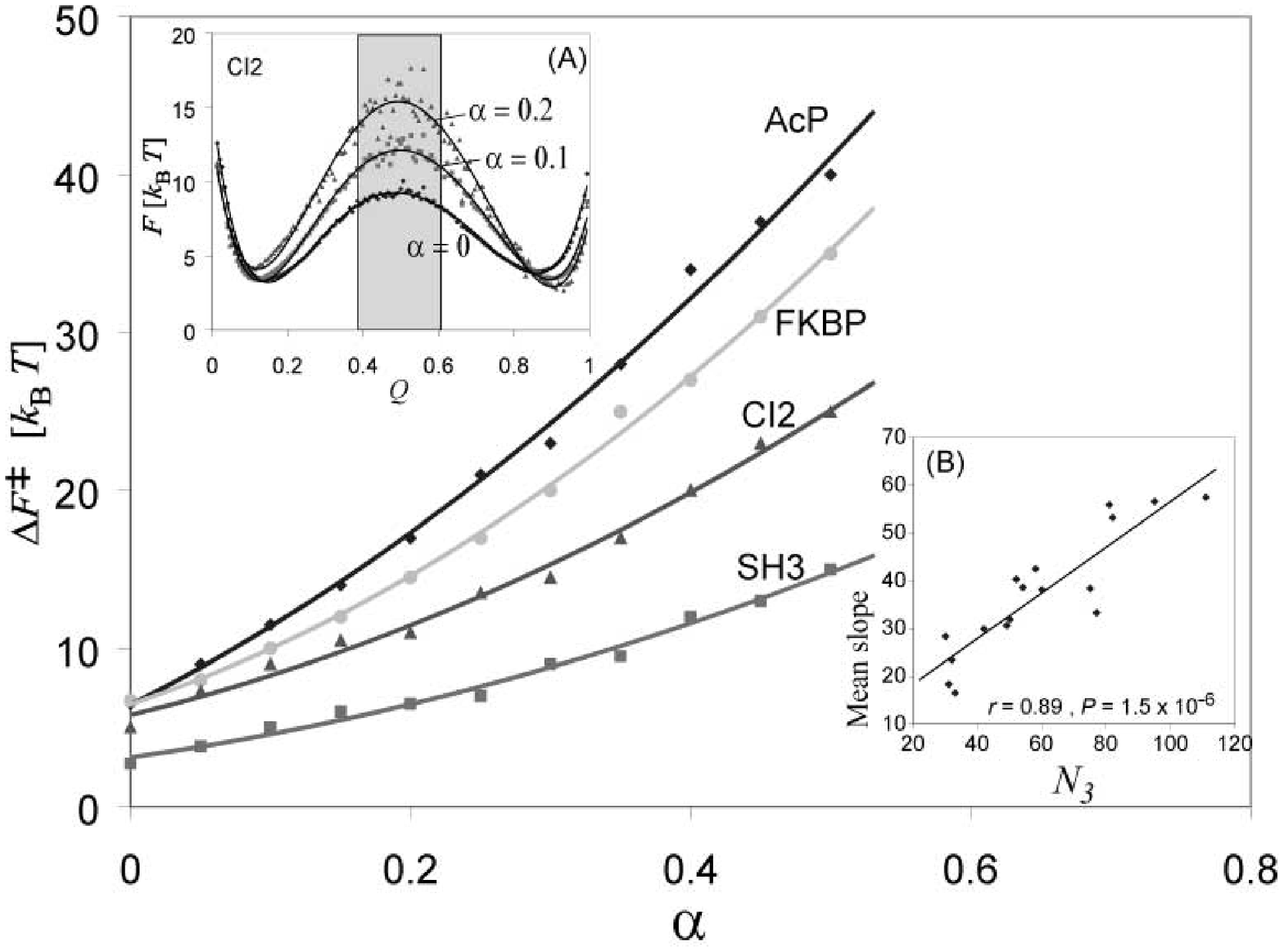}
\caption{\label{fig:df_alpha}}
\end{figure}

\vspace{1cm}
{\bf Figure:~\ref{fig:df_alpha}}\\ The folding barrier height \df increases
with increasing three-body contribution to the energy $\a$. Inset {\bf
  (A)} shows the free energy vs. the fraction 
of native contacts $Q$ for CI2, for 3 values of $\alpha$.  Main panel
shows the barrier vs. $\a$ for 4 proteins selected from
table~I.  Inset {\bf (B)}: the average slope of $\Del \Fdag$
vs. $\a$ correlates strongly with the number of $3$-body interactions
in the native state ($r=0.89$, $p=10^{-6}$).  Therefore the barriers
in the main panel increase at different rates due to differing numbers
of triples formed in the transition states of the various proteins-
more native triples typically means a larger $3$-body contribution to
the barrier. The shaded region in inset~(A) corresponds to the thermal
transition state ensemble described in the methods section. In general
this ensemble depends on $\alpha$.

\begin{figure}
\includegraphics[width=0.8\linewidth]{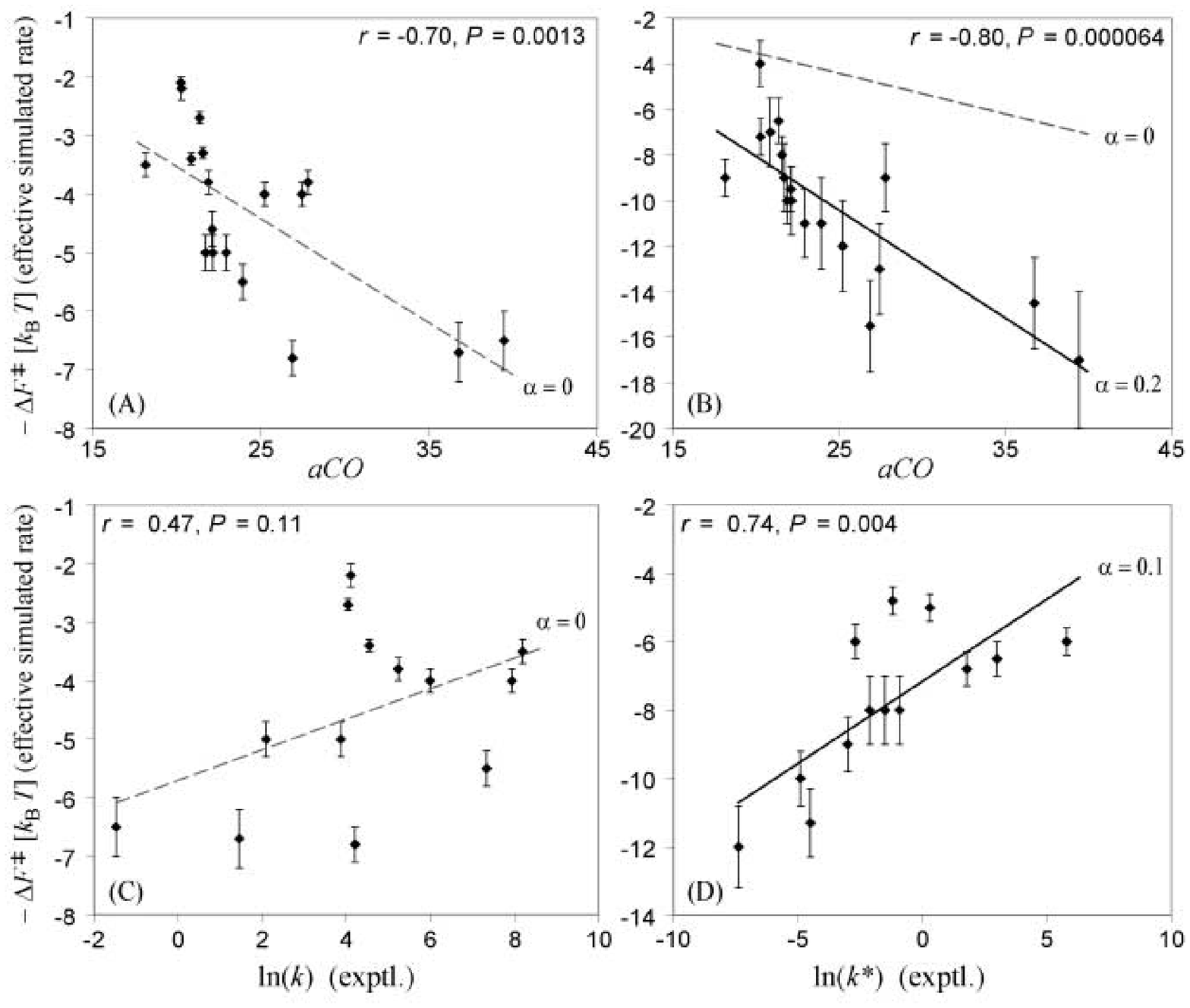}
\caption{\label{fig:df_rco}}
\end{figure}

\vspace{1cm}
{\bf Figure:~\ref{fig:df_rco}}\\ Comparison of simulated and experimental rates.
{\bf (A): } Simulated folding barriers
(effectively measuring log folding rates for 18 proteins listed in
methods) for a pair-wise interacting G\={o} model correlate well with
absolute contact order ($aCO$)~\cite{KogaN01}.  {\bf (B): } Simulated
folding barriers show an increased correlation with $aCO$, when the
fraction of native three-body energy is such that the dispersion in
effective simulated rates matches the experimental dispersion for this
data set ($\alpha = 20\%$). Rates now span $5.7$ decades, in contrast
to $2$ decades for a pure 2-Body Hamiltonian (dashed line in (B) is the
best fit line in (A)).  {\bf (C): } For $13$ of the $18$ proteins 
(see methods for a list), rate data was available for various
different denaturant concentrations. These proteins were
used for the analysis in figures~C and~D. 
Panel (C) shows that for these proteins, the simulated
effective log rates do not correlate significantly with the
experimental rate data at $25^o\, C$.  {\bf (D): } Tuning the rate
data to the transition midpoints and introducing 3-body energy in the
native state, we saw a significant increase in the correlation between
experimental and simulated rate data, with best correlation when
$\alpha = 10\%$.

\begin{figure}
\includegraphics[width=0.8\linewidth]{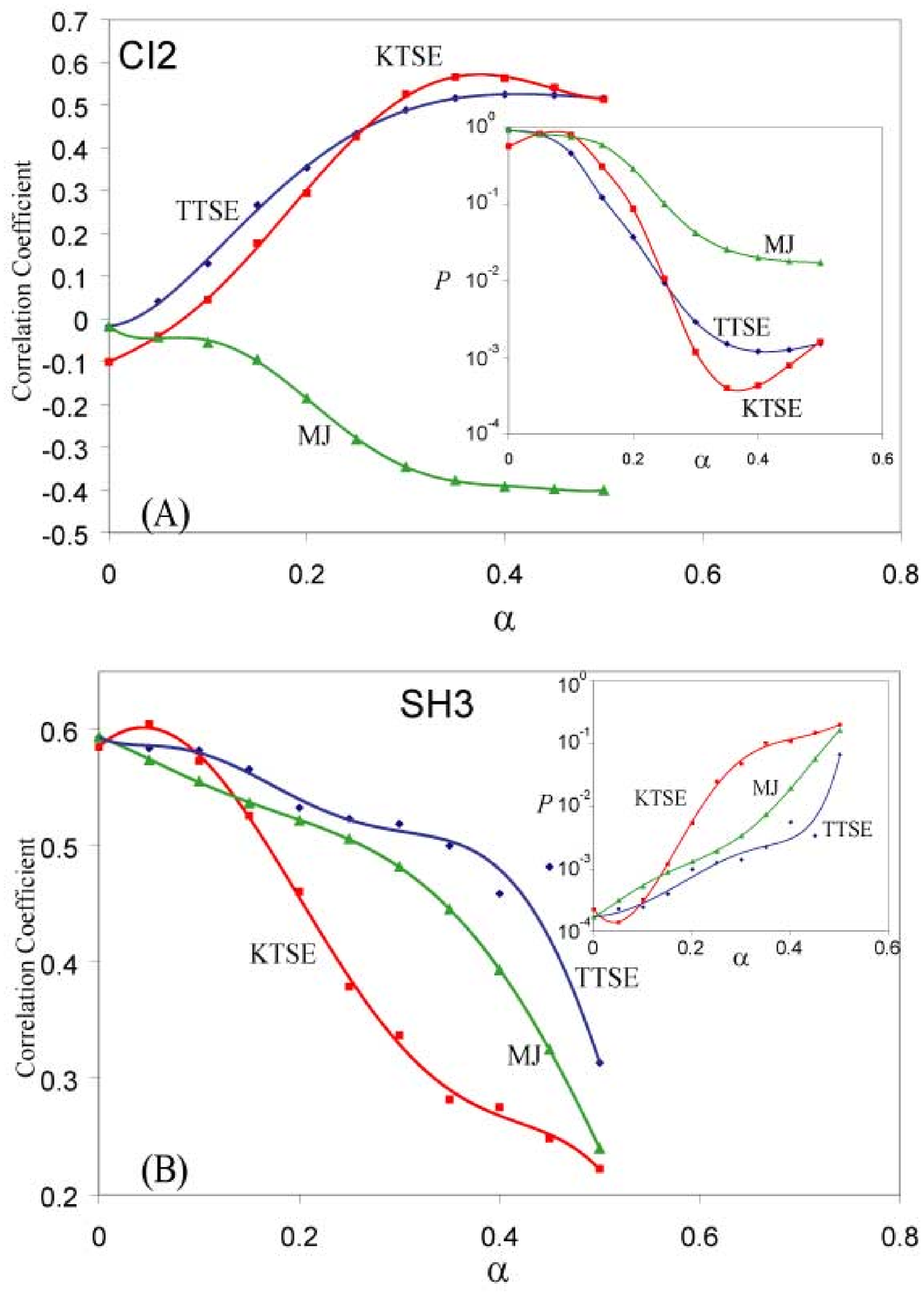}
\caption{\label{fig:phi_corr_alpha}}
\end{figure}

\vspace{1cm}
{\bf Figure:~\ref{fig:phi_corr_alpha}}\\ Comparison of the agreement of \pvs
between simulation and experiment for {\bf (A)} CI2, and {\bf (B)} src
SH3.  Green curves in A and B show the correlation coefficient and
statistical significance (insets) for \pvs derived from the thermal
transition state (TTSE) in the simulations, as the Hamiltonian was
continuously changed from a uniform G\={o} model to one with pair
interactions governed by Miyazawa Jernigan parameters (the curve
shown in inset A is the statistical significance of the
{\it anti}-correlation in the main panel) - see
equation~(\ref{eq:mj}).  No improvement was seen for CI2 or SH3 by
implementing this recipe.  Red and Blue curves show the correlation
coefficient and statistical significance between experimental and
simulated \pvs as a function of the fraction $\a$ of three-body energy
in the native state. Blue curves correspond to TTSE, Red curves-
kinetic transition state ensemble (KTSE).  For CI2 the improvement as
$\a$ is increased is dramatic, with best agreement with experiment
around $35\%$ 3-body energy. On the other hand, SH3 was exceptional in
that it showed the opposite trend, with best agreement for a purely
pair-wise interacting model for the TTSE and $\alpha=5\%$ for the
KTSE. All other proteins studied were bracketed by these two extremes-
they showed moderate components of 3-body energy, with moderate to
large increases in correlation coefficient (table~I).

\begin{figure}
\includegraphics[width=0.8\linewidth]{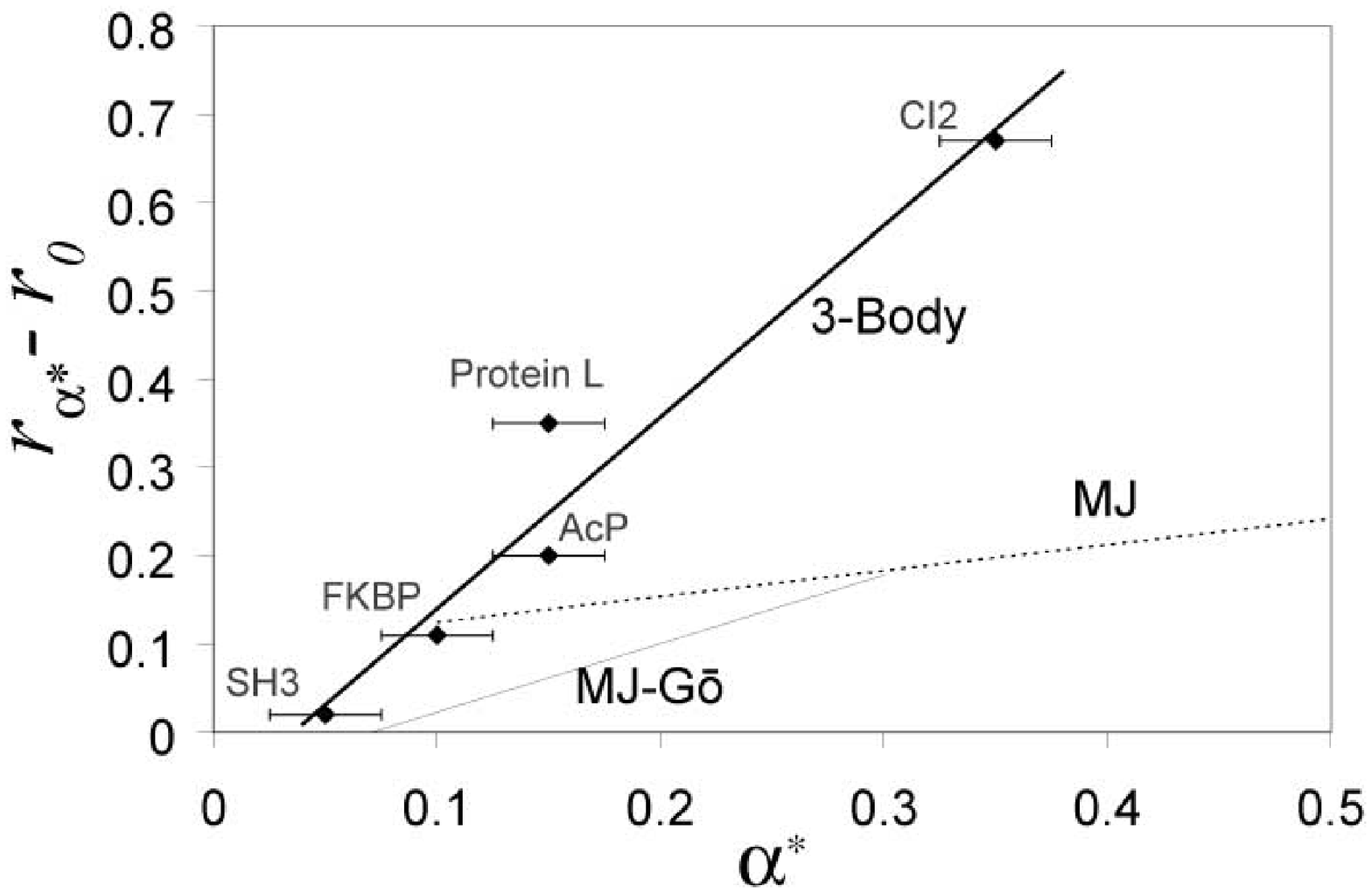}
\caption{\label{fig:drvsa}}
\end{figure}

\vspace{1cm}
{\bf Figure:~\ref{fig:drvsa}}\\ Plot of the 
largest improvement in correlation $(r_{\alpha^*}
- r_o)$ vs. the value of interpolation parameter $\alpha^*$ required
to achieve that correlation. Energy functions are interpolated toward
a 3-body G\={o} model (eq.~(\ref{eq:ENL3body})) and 2-body models with
Miyazawa-Jernigan energetic parameters (eq.~(\ref{eq:mj})). The
slope and correlation indicate the validity of the interpolation
procedure. Adding 3-body energies gives a slope of $2.2$, and $(r,P) =
(0.97 ,   0.005 )$. Adding a MJ component to the pair interaction
energies gives a slope of $0.29$ but a fit that is not statistically
significant: $(r,P) = (0.83 ,   0.38)$. Restricting the MJ component to
native interaction energies gives a statistically significant fit, 
$(r,P) = (0.956  ,  0.044)$, but with a shallow slope ($0.78$)
indicating only moderate improvement.

\begin{figure}
\includegraphics[width=0.45\linewidth]{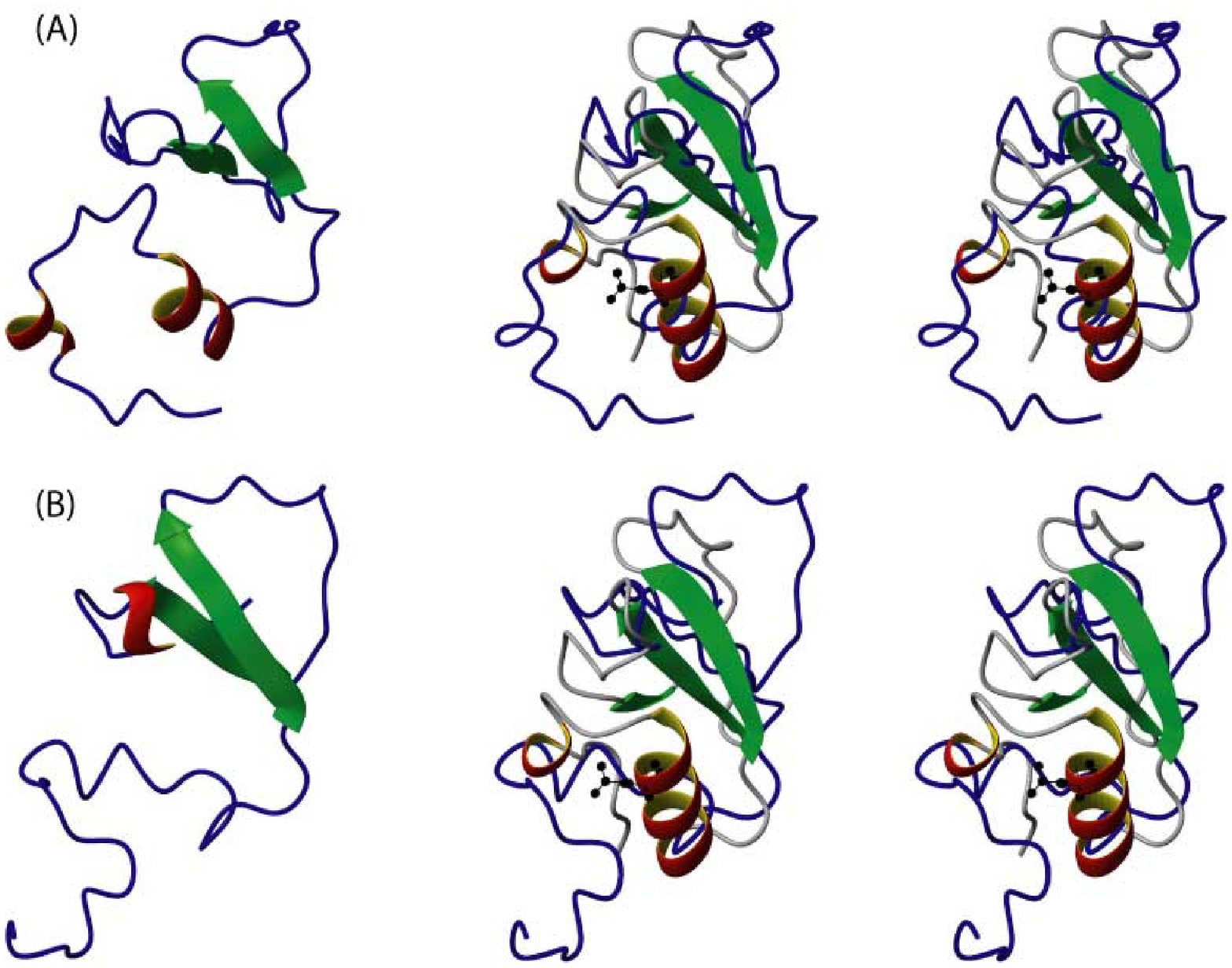}
\includegraphics[width=0.45\linewidth]{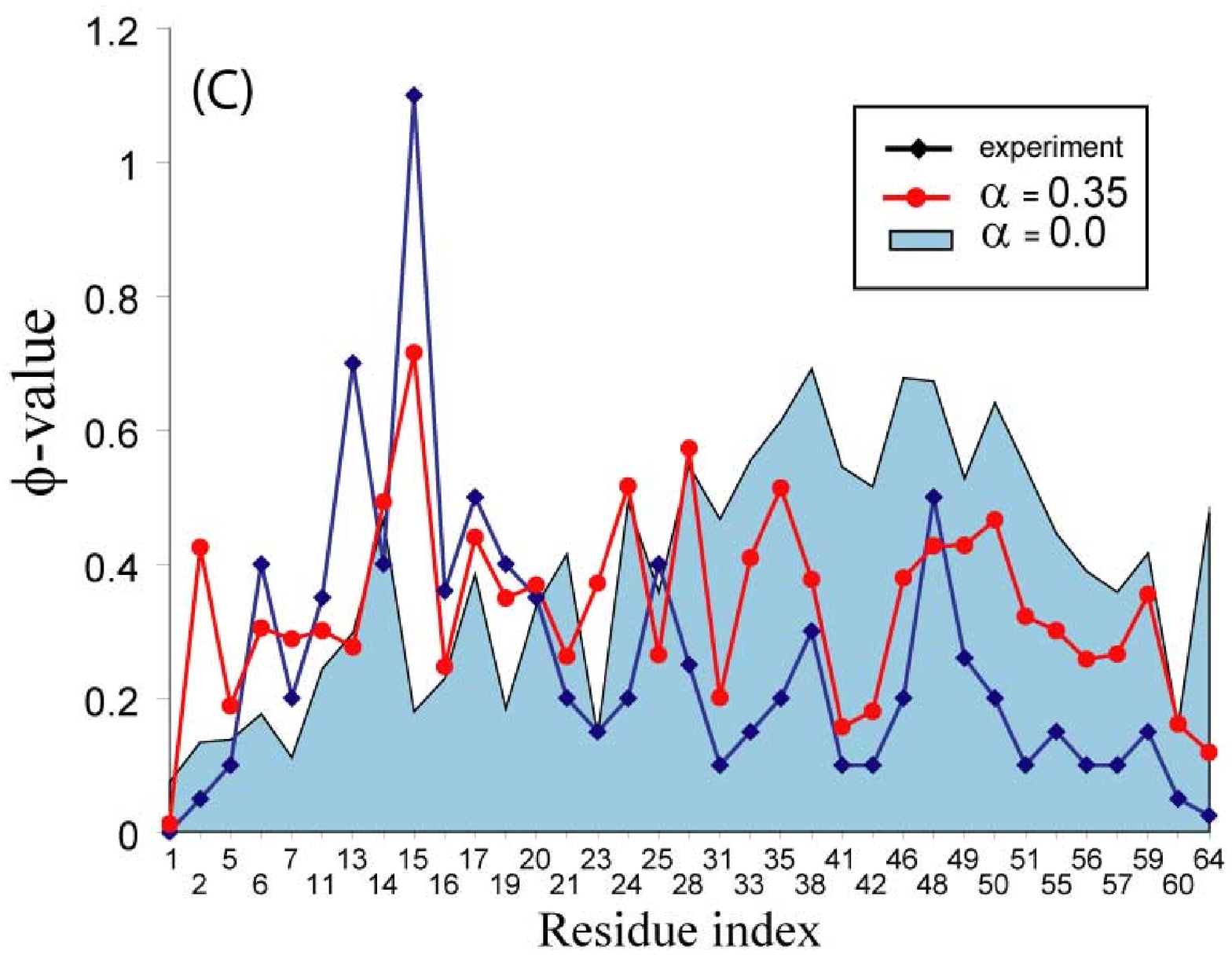}
\caption{\label{fig:TS}}
\end{figure}

\vspace{1cm}
{\bf Figure:~\ref{fig:TS}}\\ The ``most representative'' transition state
structure for the 2-body (A) and 2+3-body (B) cases of CI2, defined as
the structure having minimal Boltzmann-weighted RMSD to all other
structures in the KTSE (see text).  (left column: representation
showing secondary structure, right columns: stereographic views
superimposed on the native structure (structures generated with {\it
molmol})).  The 2-body case shows more overall secondary structure, in
particular more $\alpha$-helix, but less $\beta$-sheet.  (C): \pv
vs. residue index for CI2, for experiment (Blue), simulated pair-wise
G\={o} model (light blue background), and 2+3-body G\={o} model
(Red). The average \pvs for the various energy functions are
$\overline{\phi}^{\mbox{\tiny{(Expt)}}} = 0.25$,
$\overline{\phi}^{\mbox{\tiny{(2)}}} = 0.40$,
$\overline{\phi}^{\mbox{\tiny{(2+3)}}} = 0.33$, again confirming the
more accurate 2+3-body transition state is less structured.  It is
worth noting that native state is more stable in the experiments than
in the simulations- the native stability is
fixed at the transition midpoint
in the simulations, regardless of the value of $\alpha$.

\newpage
\newpage
.

\end{document}